\documentclass[10pt, conference]{IEEEtran}
\usepackage{cite}
\usepackage{amsmath,amssymb,amsfonts}
\usepackage{textcomp}
\usepackage{fancyvrb}

\usepackage{url}
\usepackage{listings}
\usepackage{epigraph}
\usepackage{multicol}
\usepackage{enumitem}
\usepackage[pdftex]{graphicx,xcolor}

\begin{document}

\title{How Do Programmers Express High-Level Concepts using Primitive Data Types?}

\author{\IEEEauthorblockN{1\textsuperscript{st} Yusuke Shinyama}
\IEEEauthorblockA{\textit{dept. name of organization (of Aff.)} \\
\textit{Tokyo Institute of Technology}\\
Meguro-ku, Tokyo, Japan \\
euske@sde.cs.titech.ac.jp}
\and
\IEEEauthorblockN{2\textsuperscript{nd} Yoshitaka Arahori}
\IEEEauthorblockA{\textit{dept. name of organization (of Aff.)} \\
\textit{Tokyo Institute of Technology}\\
Meguro-ku, Tokyo, Japan \\
arahori@c.titech.ac.jp}
\and
\IEEEauthorblockN{3\textsuperscript{rd} Katsuhiko Gondow}
\IEEEauthorblockA{\textit{dept. name of organization (of Aff.)} \\
\textit{Tokyo Institute of Technology}\\
Meguro-ku, Tokyo, Japan \\
gondow@cs.titech.ac.jp}
}

\maketitle

\begin{abstract}
  We investigated how programmers express high-level concepts such as
  path names and coordinates using primitive data types.  While
  relying too much on primitive data types is sometimes criticized as
  a bad smell, it is still a common practice among programmers. We
  propose a novel way to accurately identify expressions for certain
  predefined concepts by examining API calls. We defined twelve
  conceptual types used in the Java Standard API. We then obtained
  expressions for each conceptual type from 26 open source
  projects. Based on the expressions obtained, we trained a decision
  tree-based classifier. It achieved 83\% F-score for correctly
  predicting the conceptual type for a given expression.  Our result
  indicates that it is possible to infer a conceptual type from a
  source code reasonably well once enough examples are given.  The
  obtained classifier can be used for potential bug detection, test
  case generation and documentation.
\end{abstract}



\section{Introduction}

Today, the benefits of type system in programming languages are well
understood. Since a well-defined type system can prevent a programmer
from doing certain invalid operations, it helps a programmer to
achieve the correctness and safety. In a statically typed language,
proper typing also helps maintenance as it indicates the programmer's
intention.  However, defining a domain-specific type system for every
concept in a program is cumbersome. At some point, a programmer has to
rely on a more primitive data type that is closer to the runtime
environment.

\begin{figure}[htbp]
\centering
\begin{enumerate}[label=\alph*)]
\item Program with ``primitive obsession'':
\begin{Verbatim}[frame=single, fontsize=\scriptsize]
String username = getCurrentUserName();
String path = "/home/"+username+"/user.cfg";
// Unsafe path: extra check is needed!
File config = new File(path);
\end{Verbatim}
\item Equivalent well-typed program:
\begin{Verbatim}[frame=single, fontsize=\scriptsize]
User user = getCurrentUser();
Path path = Paths.get(
    user.getHomeDirectory(), "user.cfg");
// Path is guaranteed to be safe.
File config = new File(path);
\end{Verbatim}
\end{enumerate}
\caption{Primitive obsession (Java)}
\label{fig:primobs}
\end{figure}

In fact, programmers tend to use a lot of primitive data types for a
variety of purposes. In particular, string and integer are among most
commonly used data types in modern programming languages. A string
variable, for example, can be used for storing any text content, such
as user name, address and phone number. Strings are so versatile that
some programming languages only support the string data type
\cite{tcltk}. An integer is also versatile in that it can be used for
a size, counter, index, flag or other enumerable constants.

While using these primitive data types is often beneficial to
programmers, this tendency is sometimes accused as {\it primitive
  obsession} \cite{primitive-obsession} (Fig. \ref{fig:primobs}), as
it obscures the programmers' intention and poses a threat to its
safety and maintainability.  One of the major
benefits of using a well-defined abstract type system is its ability
to check the correctness of its operations.  Relying on
primitive data types means that programmers are bypassing some
of the necessary checks, resulting an unreliable or undefined behavior
of a program. For example, in most operating systems, an arbitrary
string cannot be used as a file name because a file name cannot
contain certain characters (such as ``{\tt /}'').

While primitive obsession is known to be risky, programmers often rely
on appropriate variable names and source code comments in attempt to
reduce its risk by reminding themselves its intended uses and the
domain specific constraints. It is known that programmers heavily rely
on meaningful identifier names (type/class names, function/method
names and variable/field names) to encode their intention
\cite{lawrie_06b}.

We are interested in how programmers indicate the existence of domain
specific values in source code. In this paper, we propose a novel way
to identify the expressions for several predefined concepts such as a
path name or calendar year. We applied our method to 26 well-known
open source projects and extracted the common expressions for each
concept. We then attempted to interpret the obtained results by
developing a decision tree-based classifier that infers the type of
concepts from a given expression. Our result indicates that there is a
widely used convention to express certain conceptual types in source
code. The potential applications of our technique include additional
type checking, test case generation and documentation. In the example
illustrated in Fig. \ref{fig:primobs}, one can insert an extra check to
ensure the path is correct, knowing the {\tt path} String variable
indeed specifies a file system path.

\subsection{Contribution of This Paper}
\label{subsec:contribution}

In this paper, we attempted to answer the following research
questions:

\begin{enumerate}[leftmargin=*, label={RQ.\arabic*)}]
\item What kinds of high-level concepts do programmers commonly
  use in software projects?
\item How do programmers express such concepts in source code?
\item Is it possible to accurately predict such concepts from the
  source code appearance?
\end{enumerate}

In the rest of this paper, we first define the concept of ``conceptual
types (c-types)'' in Section \ref{subsec:c-type}. We then describe how
to extract conceptual types from source code in Section
\ref{subsec:extract}. Section \ref{sec:experiments} presents the
experiment setup for obtaining conceptual type expressions and its
results. In Section \ref{sec:infer}, we analyzed the obtained
expressions for each conceptual type by constructing a decision
tree-based classifier.  Finally, we discuss our findings and the
threats to its validity in Section \ref{sec:discussions}. The related
work is described in Section \ref{sec:related}.


\section{Related Work}
\label{sec:related}

Identifying conceptual (abstract) types used in software has been an
active research topic in the field of program comprehension and
software maintenance.  O'Callahan et al. \cite{ocallahan_97} performed
type inference of a given program using static data flow analysis and
static point-to analysis. Their notion of a type is solely based on
data flow and close to an equivalence class, in that two values can
have the same type if their values can be stored into the same memory
location.  Guo et al. \cite{guo_06} took a similar approach using
dynamic analysis.  They also used the data flow of a program as a main
source of abstract type identification. Since their method does not
rely on source code, their technique could be also applied to a binary
program.  This line of research was further extended by Dash et al.
\cite{dash_18}. They combined the lexical information of a program
(variable names) with its data flow, forming the notion of ``name
flow'' that was used for clustering and discovering abstract types.
They also provided a facility to rewrite a program in such a way that
discovered types can be automatically annotated.  While it is not type
inference per se, invariant detection techniques \cite{daikon} can
also be used for type identification, as it can discern different
constraints (hence different use cases) that each variable has.

The above approaches all aimed to discover user-defined types and
constraints. One of the difficulties in these problem formulations is
that they are all somewhat subjective; there are a number of ways to
design abstract types for a particular application. The above three
approaches all used some sort of clustering technique and let the
abstract types ``emerge'' from a program. However, it is often hard to
tell if the obtained clustering was optimal for its user, as different
programs have slightly different requirements for its design.  Our
approach is different in that our conceptual types are already
well-defined by the API specification and used by many applications.
While it is not directly competing with the above three, our technique
can be used as a foundation of more advanced analysis.

Our approach is also related to the studies about program identifiers.
The importance of names in a program code has been emphasized by many
researchers and practitioners \cite{kernighan_99, mcconnel_04}.
Programmers generally prefer a long descriptive name than
single-letter variables \cite{beniamini_17}.  Poor naming can lead to
misunderstanding or confusion among programmers, which eventually
result in poor code quality \cite{avidan_17}. In some software
projects, inconsistent naming is actually considered as bugs ({\it
  naming bugs} \cite{host_09}).  Alon et al. converted source code
into word embeddings \cite{mikolov_13} that correspond to a certain
word in natural language \cite{alon_19}, which can be used for
identifiers.

In numerical or business applications, there are similar concepts to
conceptual types that are called ``dimensions''. Dimensions are
typically used for expressing physical units. Jiang et al. proposed a
way to add manual annotation of physical units to C programs and
verify their conversion to different dimensions using predefined rules
\cite{jiang_06}. Hangal et al. used a source code revision history to
check if the dimensions of each variable is consistent throughout the
development \cite{hangal_09}.


\section{Methodology}
\label{sec:methodology}

\subsection{What is Conceptual Type?}
\label{subsec:c-type}

Our basic idea is to use API specifications for capturing domain
specific values.  Well-designed API specifications usually provide a
clear definition of its inputs and outputs to each function. Since
programmers typically treat API functions as a black box, they need to
be aware of the function parameters. More specifically, they need a
precise understanding of the type of data that is being passed and how
they are going to be used. Consider the following Java example:

\begin{Verbatim}[frame=single, framesep=1em, fontsize=\scriptsize]
String x = "foo/bar.txt";
var f = new java.io.File(x);  // x is a path name.
\end{Verbatim}

In the above snippet, the first argument of the {\tt java.io.File}
constructor is supposed to be of {\tt String} type, according to the
Java API specification. However, the programmer has to be aware that
it has a more strict requirement than {\it just} a string because it
has to be a path name.  Therefore, the programmer is responsible to
make sure that the value of {\tt x} is not just a string but it meets
the requirements of a valid path name (such as not containing invalid
characters). In this sense, the first argument of the {\tt File}
constructor requires a more specific data type than ones that are
provided by the programming language.

Conceptually, API entry points presents a clear boundary that
translates primitive data types such as string to a more
specific domain. In contrast to data types provided by a programming
language, we call these data types a ``{\it conceptual type}'' (or
``{\it c-type}'' in short).

We identified c-types that frequently appear in the Java Standard API
\cite{java-api}. The principles we used in choosing these c-types are
the following:

\begin{enumerate}
\item It has a clearly defined concept that is well understood by most programmers.
\item It is distinct enough that people do not mix up with other concepts.
\item It is widely used in a variety of applications.
\end{enumerate}

Table \ref{tab:c-types} lists the 12 c-types we chose. The domain of
these c-types can be divided into four different sections: file I/O,
networking, GUI (Java AWT) and date/time handling. These domains
are general enough that can be used in a variety of software projects.

Note that XCOORD (X coordinate) and YCOORD (Y coordinate) are treated
as a separate type, as well as WIDTH and HEIGHT. These types could be
merged into one, as they all represent a distance or length in a
graphical device.  However, programmers rarely treat these values
interchangeably\footnote{ While expressions like {\tt Point(x, x)} or
{\tt x+width*2} might be used in some programs, we can hardly imagine
a GUI program where operations like {\tt x+y} or {\tt Point(y, x)} are
meaningful.}.  Following the above principle 1, we consider them
different c-types.

After choosing the c-types, we identified the methods that take one or
more of the defined types as arguments. Table \ref{tab:typemethods}
and Fig. \ref{fig:methods} show the number of the method arguments
selected and the excerpt of these methods, respectively. In total, we
selected 218 methods including overlaps.  Note that some methods take
multiple c-types as its arguments at once (such as WIDTH and HEIGHT).

\begin{table}[htbp]
\caption{Conceptual Types (C-Types)}
\begin{center}
\begin{tabular}{|l|l|l|}
\hline
C-Type & Actual Type & Description \\
\hline
PATH   & {\tt String} & Path name         \\
URL    & {\tt String} & URL/URI           \\
SQL    & {\tt String} & SQL statement     \\
HOST   & {\tt String} & Host name         \\
PORT   & {\tt int}    & Port number       \\
XCOORD & {\tt int}    & X coordinate (for GUI)  \\
YCOORD & {\tt int}    & Y coordinate (for GUI)  \\
WIDTH  & {\tt int}    & Width (for GUI)   \\
HEIGHT & {\tt int}    & Height (for GUI)  \\
YEAR   & {\tt int}    & Year              \\
MONTH  & {\tt int}    & Month             \\
DAY    & {\tt int}    & Day of month      \\
\hline
\end{tabular}
\end{center}
\label{tab:c-types}
\end{table}

\begin{table}[htbp]
\caption{Number of Methods for Each C-Type}
\begin{center}
\begin{tabular}{|l|r|}
\hline
C-Type & \# Methods \\
\hline
PATH   & 14 \\
URL    &  4 \\
SQL    & 10 \\
HOST   & 17 \\
PORT   & 25 \\
XCOORD & 25 \\
YCOORD & 25 \\
WIDTH  & 24 \\
HEIGHT & 24 \\
YEAR   & 18 \\
MONTH  & 14 \\
DAY    & 18 \\
\hline
Total & 218 \\
\hline
\end{tabular}
\end{center}
\label{tab:typemethods}
\end{table}

\begin{figure}[htbp]
\begin{center}
\fbox{\parbox{0.9 \linewidth} {\scriptsize \begin{itemize}
    \item {\tt new java.io.File(\underline{PATH})}
    \item {\tt new java.net.URI(\underline{URL})}
    \item {\tt java.sql.Statement.execute(\underline{SQL})}
    \item {\tt java.net.InetAddress.getByName(\underline{HOST})}
    \item {\tt new java.net.Socket(\underline{HOST}, \underline{PORT})}
    \item {\tt new java.awt.Point(\underline{XCOORD}, \underline{YCOORD})}
    \item {\tt new java.awt.Dimension(\underline{WIDTH}, \underline{HEIGHT})}
    \item {\tt new java.util.Date(\underline{YEAR}, \underline{MONTH}, \underline{DAY})}
    \item {\tt java.util.Date.setYear(\underline{YEAR})}
    \item {\tt java.time.LocalDate.of(\underline{YEAR}, \underline{MONTH}, \underline{DAY})}
    \item ...
\end{itemize} }}
\end{center}
\caption{Excerpt of Methods used for Identifying C-Types}
\label{fig:methods}
\end{figure}

\subsection{Extracting Conceptual Type Expressions}
\label{subsec:extract}

With the list of methods that define c-types, we scan the source code
and identify all the calls for the selected methods or
constructors. Each method call has one or more arguments that specify
a predefined c-type. We then extract the expressions for each
argument as a {\it c-type expression}.

In this paper, we use Java as our target language. First, we identify
all the methods (including overloaded methods) and assign a unique
identifier to each.  We keep a list of method identifiers (the names
and signatures) that we selected and check if each method
call can match those identifiers.

In a case of virtual method call (dynamic dispatching), there are
multiple method implementations that has the same signature. Note that
we are only interested in the arguments of each method call; we do not
need to know which method is actually invoked.  When a method call can
potentially invoke multiple implementations, we collect its arguments
if one of its possible destinations is defined in our method list.

\subsection{Implementation}
\label{subsec:impl}

We implemented a static analyzer for Java source code.  The analyzer
takes the following steps for the given set of files:

\begin{enumerate}
\item Parse all the source codes.
  We used Eclipse JDT \cite{eclipse-jdt} for the Java parser.
\item Enumerate all the classes and name spaces defined in the
  target source code. We maintain a hierarchical symbol table
  for registering Java packages.
\item Process {\tt import} statements in each file to resolve the
  references to external classes.
\item Scan all the method signatures and assign a unique identifier
  to each method. For example, a method which has a signature:
\begin{Verbatim}[frame=single, framesep=1em, fontsize=\scriptsize]
package foo.bar;
class Config {
    int findString(String s[], int i)
}
\end{Verbatim}
can be encoded as a unique identifier:
\begin{center} \scriptsize
\verb+foo.bar.Config.findString([LString;I)I+
\end{center}
\item In addition to source codes, compiled Java class files
  and {\tt jar} files are also scanned and its method signatures
  are collected.
\item Construct a symbol table that includes all the variables and
  field names defined in each method. The symbol table has
  mappings from a variable (field) name to its data type.
  The symbol table is used for method resolution in the next step.
\item For every method or constructor call,
  find the most precise method that matches
  the calling signature.
\item If the callee method is one of the selected methods (i.e. its
  method identifier is in our list), extract the corresponding
  arguments that specify one of the predefined c-types.
\end{enumerate}

Note that Step 3 above typically requires complete type information
for imported classes. In our case, however, since we only need to
identify the calls of the methods that we selected, not all the
references need to be resolved. Since all the methods we chose are
included in the Java Standard API, we simply ignored unresolved method
calls. This allows us to process a variety of Java projects without
needing its dependencies.


\section{Experiments}
\label{sec:experiments}

We extracted c-type expressions from 26 open source projects.  First we
listed top 1,000 Java projects in the number of stars in GitHub. We
then performed string search through their source code and selected
ones that uses one or more of the Java APIs listed in Table
\ref{tab:typemethods}. We chose projects of a variety of sizes.  The
size of each project ranges from 1.8mLoC to 3kLoC. Table
\ref{tab:projects} shows the projects and their sizes.

We used a standard PC (Intel Xeon 2.2GHz, 40 core, 64G bytes memory,
running Arch Linux) for running our experiment.  Extracting method
calls and c-type expressions for the all 26 projects took less than 2
hours in total.

\begin{table}[htbp]
\caption{Projects and Sizes (LoC was counted with \cite{sloccount})}
\begin{center}
\begin{tabular}{|l|l|r|}
\hline
Project & Description & LoC \\
\hline
hadoop 3.3.1      & distributed computation & 1,789k \\
ghidra 10.0       & binary analyzer         & 1,588k \\
ignite 2.10.0     & distributed database    & 1,165k \\
jetty 11.0.5      & web container           &   441k \\
kafka 2.7.1       & stream processing       &   384k \\
tomcat 8.5.68     & web server              &   349k \\
jitsi 2.10        & video conference        &   327k \\
binnavi 6.1.0     & binary analyzer         &   309k \\
netty 4.1.65      & network library         &   303k \\
libgdx 1.10.0     & game framework          &   272k \\
alluxio 2.5.0-3   & data orchestration      &   228k \\
plantuml 1.2021.7 & UML generator           &   210k \\
grpc 1.38.1       & RPC framework           &   195k \\
jenkins 2.299     & automation              &   177k \\
jmeter 5.4.1      & network analyzer        &   145k \\
jedit 5.6.0       & text editor             &   125k \\
gephi 0.9.2       & graph visualizer        &   120k \\
zookeeper 3.7.0   & distributed computation &   114k \\
selenium 3.141.59 & browser automation      &    91k \\
okhttp 4.9.1      & HTTP client             &    36k \\
jhotdraw 7.0.6    & graph drawing           &    32k \\
arduino 1.8.15    & development environment &    27k \\
gson 2.8.7        & serialization framework &    25k \\
websocket 1.5.2   & network framework       &    15k \\
picasso 2.8       & image processing        &     9k \\
jpacman           & action game             &     3k \\
\hline
\multicolumn{2}{|l|}{Total} & 8,480k \\
\hline
\end{tabular}
\end{center}
\label{tab:projects}
\end{table}

Table \ref{tab:catcount} shows the number of extracted c-type
expressions for each project. The ``OTHER'' column shows not an actual
c-type, but the number of expressions that are passed in arguments that
does not specify any predefined c-type. For example, some API method
takes a path name and an extra boolean flag as arguments.  Since this
extra argument does not specify any predefined c-type, we count them as
OTHER. The OTHER expressions are later used for training the decision
tree algorithm and measuring its performance\footnote{ In theory, all
the arguments of all the method calls that we are not interested in
should be counted as the OTHER type.  For practical reasons, however,
we ignored method calls that clearly have nothing to do with c-types.
}.

\begin{table*}[htbp]
\caption{Extracted C-Type Expressions by Project}
\begin{center}
\begin{tabular}{|l|r|r|r|r|r|r|r|r|r|r|r|r|r|r|r|}
\hline
Project & LoC & {\tiny PATH} & {\tiny URL} & {\tiny SQL} & {\tiny HOST} & {\tiny PORT} & {\tiny XCOORD} & {\tiny YCOORD} & {\tiny WIDTH} & {\tiny HEIGHT} & {\tiny YEAR} & {\tiny MONTH} & {\tiny DAY} & {\tiny OTHER} & All \\
\hline
   alluxio &  228k &  72 &  12 &   0 &  22 &  26 &   0 &   0 &   0 &   0 &  15 &  15 &  15 &  51 &  228 \\
   arduino &   27k &  39 &   5 &   0 &  12 &   9 &   6 &   6 &  42 &  42 &   0 &   0 &   0 &   1 &  162 \\
   binnavi &  309k &  42 &   7 &   1 &   2 &   1 &  23 &  23 &  67 &  67 &   1 &   0 &   0 &   1 &  235 \\
     gephi &  120k &   5 &   0 &   2 &   0 &   0 &  28 &  28 &  66 &  66 &   0 &   0 &   0 &   0 &  195 \\
    ghidra &1,588k & 369 &  25 &   0 &  10 &   8 & 320 & 320 & 511 & 511 &  13 &  13 &  13 &  13 & 2,126 \\
      grpc &  195k &  33 &  16 &   0 &  68 &  71 &   0 &   0 &   0 &   0 &   0 &   0 &   0 &  75 &  263 \\
      gson &   25k &   0 &   4 &   0 &   2 &   0 &   0 &   0 &   0 &   0 &   6 &   6 &   6 &   7 &   31 \\
    hadoop &1,789k & 978 & 634 &   9 & 288 & 259 &   0 &   0 &   0 &   0 &   2 &   2 &   2 & 124 & 2,298 \\
    ignite &1,165k & 168 &  85 & 666 & 106 & 111 &   0 &   0 &   0 &   0 &  12 &  12 &  12 & 101 & 1,273 \\
     jedit &  125k & 130 &  19 &   0 &   3 &   3 &  50 &  50 & 112 & 112 &   0 &   0 &   0 &   3 &  482 \\
   jenkins &  117k &  82 &  28 &   0 &   6 &   6 &   1 &   1 &   1 &   1 & 102 & 102 & 102 & 237 &  669 \\
     jetty &  441k &  72 & 216 &   9 & 163 & 104 &   0 &   0 &   0 &   0 &   0 &   0 &   0 &  37 &  601 \\
  jhotdraw &   32k &   6 &   5 &   0 &   1 &   1 &  96 &  96 &  50 &  50 &   0 &   0 &   0 &   1 &  306 \\
     jitsi &  327k &  22 &  18 &   1 &   8 &  31 &  78 &  78 & 234 & 234 &   0 &   0 &   0 &  30 &  734 \\
    jmeter &  145k & 112 &  62 &   2 &  31 &  28 &   7 &   7 &  62 &  62 &   0 &   0 &   0 &  28 &  401 \\
   jpacman &    3k &   0 &   0 &   0 &   0 &   0 &   0 &   0 &   1 &   1 &   0 &   0 &   0 &   0 &    2 \\
     kafka &  384k &  37 &   1 &   0 &  85 &  72 &   0 &   0 &   0 &   0 &  14 &  14 &  14 &  44 &  281 \\
    libgdx &  272k &  83 &   7 &   0 &   4 &   5 &   7 &   7 &  36 &  36 &   0 &   0 &   0 &   0 &  185 \\
     netty &  303k &  38 &  24 &   0 &  54 & 130 &   0 &   0 &   0 &   0 &   1 &   1 &   1 & 117 &  366 \\
    okhttp &   36k &   2 &  24 &   0 &   6 &   7 &   0 &   0 &   0 &   0 &   0 &   0 &   0 &   3 &   42 \\
   picasso &    9k &   1 &   0 &   0 &   0 &   0 &   0 &   0 &   0 &   0 &   0 &   0 &   0 &   0 &    1 \\
  plantuml &  210k &  29 &   4 &   0 &   5 &  11 &   4 &   4 &  11 &  11 &   2 &   2 &   2 &   2 &   87 \\
  selenium &   91k &  44 &  66 &   0 &  15 &  12 &   1 &   1 &   0 &   0 &   1 &   1 &   1 &  21 &  163 \\
    tomcat &  349k & 207 &  64 &  22 &  38 &  52 &   0 &   0 &  10 &  10 &   0 &   0 &   0 &  47 &  450 \\
 websocket &   15k &   9 &  44 &   0 &   5 &  43 &   0 &   0 &   1 &   1 &   0 &   0 &   0 &   1 &  104 \\
 zookeeper &  114k &  88 &   4 &   0 & 126 & 192 &   0 &   0 &   1 &   1 &   0 &   0 &   0 &  47 &  459 \\
\hline
Total &8,480k & 2,668 & 1,374 & 712 & 1,060 & 1,182 & 621 & 621 & 1,205 & 1,205 & 169 & 168 & 168 & 991 & 12,144 \\
\hline
\end{tabular}
\end{center}
\label{tab:catcount}
\end{table*}


We collected frequently used expressions in each project.  Table
\ref{tab:topexprs} shows the most frequent expressions for four c-types
(PATH, URL, XCOORD and WIDTH) in each project.  Constant expressions
such as {\tt "localhost"} are excluded. While shorter and more common
expressions are relatively straightforward, a long expression with
multiple operators can be complex for programmers. Table
\ref{tab:complexity} shows the length of expressions for each c-type,
in the number of components included in each expression\footnote{ Note
that field access ({\tt a.b}) and method call ({\tt a.b()}) are
considered as two components instead of one.}. Table \ref{tab:opexprs}
shows compound expressions that include binary operators (such as {\tt
  +} or {\tt *}).

We also obtained frequently used words for each c-type by using the
word segmentation algorithm shown in Section \ref{subsec:wordseg}.
The results are shown in Table \ref{tab:topwords}.

\begin{table}[htbp]
\caption{Top Expressions for PATH, URL, XCOORD and WIDTH C-Types (Constants Excluded)}
\begin{center}
\begin{tabular}{|l|l|}
\hline
 PATH & Top Expressions \\
\hline
 alluxio & {\tt \scriptsize path, mLocalUfsPath+ufsBase, base} \\
 arduino & {\tt \scriptsize path, PreferencesData.get("runtime.ide.path")} \\
 binnavi & {\tt \scriptsize filename, directory, pathname} \\
 gephi & {\tt \scriptsize System.getProperty("netbeans.user")} \\
 ghidra & {\tt \scriptsize getTestDirectoryPath(), path, filename} \\
 grpc & {\tt \scriptsize uri.getPath()} \\
 hadoop & {\tt \scriptsize GenericTestUtils.getRandomizedTempPath()} \\
 ignite & {\tt \scriptsize path, U.defaultWorkDirectory(), fileName} \\
 jedit & {\tt \scriptsize path, dir, directory} \\
 jenkins & {\tt \scriptsize System.getProperty("user.home"), war} \\
 jetty & {\tt \scriptsize file.getParent()} \\
 jhotdraw & {\tt \scriptsize prefs.get("projectFile", home)} \\
 jitsi & {\tt \scriptsize path, localPath} \\
 jmeter & {\tt \scriptsize filename, path, file} \\
 kafka & {\tt \scriptsize storeDirectoryPath, argument} \\
 libgdx & {\tt \scriptsize name, sourcePath, imagePath.replace('\textbackslash\textbackslash','/')} \\
 netty & {\tt \scriptsize getClass().getResource("test.crt").getFile()} \\
 plantuml & {\tt \scriptsize filename, newName} \\
 selenium & {\tt \scriptsize System.getProperty("java.io.tmpdir"), logName} \\
 tomcat & {\tt \scriptsize pathname, path, docBase} \\
 zookeeper & {\tt \scriptsize path, KerberosTestUtils.getKeytabFile()} \\
\hline
 URL & Top Expressions \\
\hline
 alluxio & {\tt \scriptsize journalDirectory, folder, inputDir} \\
 arduino & {\tt \scriptsize contribution.getUrl(), packageIndexURLString} \\
 binnavi & {\tt \scriptsize url, urlString} \\
 ghidra & {\tt \scriptsize ref, getAbsolutePath(), url.toExternalForm()} \\
 grpc & {\tt \scriptsize target, TARGET, oobTarget} \\
 gson & {\tt \scriptsize nextString, urlValue, uriValue} \\
 hadoop & {\tt \scriptsize uri, url, s} \\
 ignite & {\tt \scriptsize GridTestProperties.getProperty("p2p.uri.cls")} \\
 jedit & {\tt \scriptsize path, str, fileIcon} \\
 jenkins & {\tt \scriptsize url, site.getData().core.url, plugin.url} \\
 jetty & {\tt \scriptsize uri, inputUrl.toString(), s} \\
 jitsi & {\tt \scriptsize url, imagePath, sourceString} \\
 jmeter & {\tt \scriptsize url, LOCAL\_HOST, requestPath} \\
 kafka & {\tt \scriptsize config.getString(METRICS\_URL\_CONFIG)} \\
 libgdx & {\tt \scriptsize url, URI, httpRequest.getUrl()+queryString} \\
 netty & {\tt \scriptsize URL, request.uri(), server} \\
 selenium & {\tt \scriptsize url, baseUrl, (String)raw.get("uri")} \\
 tomcat & {\tt \scriptsize url, location, path} \\
 websocket & {\tt \scriptsize uriField.getText(), uriinput.getText()} \\
 zookeeper & {\tt \scriptsize urlStr} \\
\hline
 XCOORD & Top Expressions \\
\hline
 arduino & {\tt \scriptsize noLeft, cancelLeft} \\
 binnavi & {\tt \scriptsize x, m\_x} \\
 gephi & {\tt \scriptsize currentMouseX, x, bounds.x} \\
 ghidra & {\tt \scriptsize x, center.x+deltaX, filterPanelBounds.x} \\
 jedit & {\tt \scriptsize x, event.getX(), leftButtonWidth+leftWidth} \\
 jhotdraw & {\tt \scriptsize evt.getX(), x, e.getX()} \\
 jitsi & {\tt \scriptsize x, button.getX(), dx} \\
 jmeter & {\tt \scriptsize graphPanel.getLocation().x, cellRect.x, x} \\
 libgdx & {\tt \scriptsize upButtonX, getWidth()-buttonSize.width-5, x} \\
 plantuml & {\tt \scriptsize e.getX()} \\
\hline
 WIDTH & Top Expressions \\
\hline
 arduino & {\tt \scriptsize width, imageW, Preferences.BUTTON\_WIDTH} \\
 binnavi & {\tt \scriptsize COLORPANEL\_WIDTH, TEXTFIELD\_WIDTH, width} \\
 gephi & {\tt \scriptsize w, constraintWidth, DEPTH} \\
 ghidra & {\tt \scriptsize width, center.width, filterPanelBounds.width} \\
 jedit & {\tt \scriptsize width, buttonSize.width, colWidth} \\
 jhotdraw & {\tt \scriptsize frameWidth, r.width, bounds.width} \\
 jitsi & {\tt \scriptsize MAX\_MSG\_PANE\_WIDTH, WIDTH, width} \\
 jmeter & {\tt \scriptsize graphPanel.width} \\
 libgdx & {\tt \scriptsize width, buttonSize.width} \\
 plantuml & {\tt \scriptsize newWidth} \\
 tomcat & {\tt \scriptsize WIDTH} \\
\hline
\end{tabular}
\end{center}
\label{tab:topexprs}
\end{table}

\begin{table}[htbp]
\caption{Expression Length (Number of Components)}
\begin{center}
\begin{tabular}{|l|r|r|r|r|r|r|r|r|r|}
\hline
C-Type & $n=1$ & $n=2$ & $n=3$ & $n=4$ & $n=5$ & $n=6$ & $n\ge7$ \\
\hline
PATH   & 49.6\% & 22.8\% & 7.0\% & 6.3\% & 4.8\% & 2.2\% & 7.3\% \\
URL    & 31.6\% & 18.5\% & 13.7\% & 13.5\% & 10.2\% & 5.8\% & 6.8\% \\
SQL    & 47.5\% & 12.1\% & 8.7\% & 3.1\% & 4.4\% & 5.5\% & 18.8\% \\
HOST   & 59.2\% & 11.5\% & 3.0\% & 22.1\% & 2.0\% & 0.1\% & 2.1\% \\
PORT   & 68.4\% & 27.5\% & 2.1\% & 1.2\% & 0.3\% & 0.4\% & 0.0\% \\
XCOORD & 54.1\% & 24.6\% & 9.8\% & 6.4\% & 1.6\% & 2.1\% & 1.3\% \\
YCOORD & 52.5\% & 22.4\% & 10.1\% & 9.0\% & 2.6\% & 1.9\% & 1.4\% \\
WIDTH  & 71.0\% & 15.2\% & 6.3\% & 2.5\% & 1.5\% & 1.8\% & 1.7\% \\
HEIGHT & 71.4\% & 15.4\% & 6.4\% & 3.1\% & 1.5\% & 1.3\% & 1.0\% \\
YEAR   & 96.4\% & 2.4\% & 1.2\% & 0.0\% & 0.0\% & 0.0\% & 0.0\% \\
MONTH  & 79.8\% & 19.6\% & 0.6\% & 0.0\% & 0.0\% & 0.0\% & 0.0\% \\
DAY    & 99.4\% & 0.0\% & 0.6\% & 0.0\% & 0.0\% & 0.0\% & 0.0\% \\
\hline
\end{tabular}
\end{center}
\label{tab:complexity}
\end{table}

\begin{table*}[htbp]
\caption{Compound Expressions with Operators}
\begin{center}
\begin{tabular}{|l|p{0.7\linewidth}|}
\hline
 C-Type & Expressions \\
\hline
 PATH &
 {\tt \scriptsize mLocalUfsPath + ufsBase} \newline
 {\tt \scriptsize selectedFile.getAbsolutePath() + PREFERENCES\_FILE\_EXTENSION} \newline
 {\tt \scriptsize dir.getPath() + DIR\_FAILURE\_SUFFIX} \newline
 {\tt \scriptsize U.defaultWorkDirectory() + separatorChar + DEFAULT\_TARGET\_FOLDER + separatorChar}
 \\
\hline
URL &
 {\tt \scriptsize url.toExternalForm().substring(GhidraURL.PROTOCOL.length() + 1)} \newline
 {\tt \scriptsize str + KMSRESTConstants.SERVICE\_VERSION + "/"} \newline
 {\tt \scriptsize newOrigin(getScheme(),getHost(),getPort()).asString() + path} \newline
 {\tt \scriptsize base + configFile}
 \\
\hline
 XCOORD &
 {\tt \scriptsize center.x + center.width} \newline
 {\tt \scriptsize leftButtonWidth + leftWidth} \newline
 {\tt \scriptsize evt.getX() - getInsets().left} \newline
 {\tt \scriptsize prefs.getInt(name+".x", 0)}
 \\
\hline
 WIDTH &
 {\tt \scriptsize Math.max(contentWidth, menuWidth) + insets.left + insets.right} \newline
 {\tt \scriptsize TITLE\_X\_OFFSET + titlePreferredSize.width} \newline
 {\tt \scriptsize width + insets.left + insets.right + 2} \newline
 {\tt \scriptsize (int)(bounds.getWidth() * percent)}
 \\
\hline
\end{tabular}
\end{center}
\label{tab:opexprs}
\end{table*}

\begin{table}[htbp]
\caption{Top Words used in C-Type Expressions}
\begin{center}
\begin{tabular}{|l|l|}
\hline
 C-Type & Top words (\# Projects) \\
\hline
 PATH & {\tt get} (21), {\tt path} (21), {\tt file} (20) \\
 URL & {\tt url} (19), {\tt get} (18), {\tt string} (18) \\
 SQL & {\tt get} (6), {\tt query} (5), {\tt create} (3) \\
 HOST & {\tt host} (21), {\tt get} (17), {\tt address} (17) \\
 PORT & {\tt port} (22), {\tt get} (18), {\tt local} (10) \\
 XCOORD & {\tt width} (9), {\tt x} (9), {\tt get} (9) \\
 YCOORD & {\tt height} (9), {\tt y} (9), {\tt get} (8) \\
 WIDTH & {\tt width} (13), {\tt get} (11), {\tt size} (10) \\
 HEIGHT & {\tt height} (12), {\tt get} (11), {\tt size} (10) \\
 YEAR & {\tt year} (4), {\tt get} (2), {\tt int} (2) \\
 MONTH & {\tt january} (3), {\tt month} (3), {\tt december} (3) \\
 DAY & {\tt day} (3), {\tt int} (2), {\tt parse} (2) \\
\hline
\end{tabular}
\end{center}
\label{tab:topwords}
\end{table}


\section{Inferring C-Types by Expressions}
\label{sec:infer}

In this section, we describe our attempt to develop a decision
tree-based classifier that predicts the c-type from a given expression.
Since the expressions obtained for each c-type contain several words
that are commonly used across many projects, we expected that we could
construct a relatively straightforward model (if any) to infer the
c-type of a given expression.

A decision tree is a relatively simple machine learning model that is
equivalent to a sequence of if-then statements.  It is efficient and
suitable for handling discrete values such as symbols or words. One of
the major advantages of a decision tree is that it is human
readable. We used a ID3 algorithm \cite{quinlan:93} to construct a
decision tree.

In the rest of this section, we first describe how to decompose an
expression to a set of features used for inferring conceptual
types. Our classifier uses both lexical and data flow-centric
information of an expression.  Then we describe a word segmentation
algorithm used in feature extraction. A word segmentation is needed to
split identifiers that are made up with multiple words (such as {\tt
  getPath}). We then show its predictive performance and an excerpt of
obtained rules.

\subsection{Converting Expression into Features}
\label{subsec:features}

In this experiment, a c-type is specified by an argument in a method
call. Each argument is a Java expression that consists of the
following terms: Variable (field) accesses, method calls and
constants. To use a decision tree classifier, the syntax tree of each
expression needs to be converted as discrete features.

Our basic idea is to focus on each identifier in an expression in the
order of significance.  When an expression consists of only one
variable reference (such as {\tt path}), we call this variable a {\it
  primary identifier}.  When an expression consists of two references
where one variable belongs to another (such as {\tt a.b}), we choose
the most significant reference ({\tt b}) as a primary identifier as
the other ({\tt a}) as a {\it secondary identifier}.  This strategy
can be formalized by using the idea of data dependency graph (data
flow graph) which has been commonly used in compiler optimization
\cite{ferrante_87}.

\begin{figure}[htbp]
\centering
\fbox{\includegraphics[width=0.7 \linewidth]{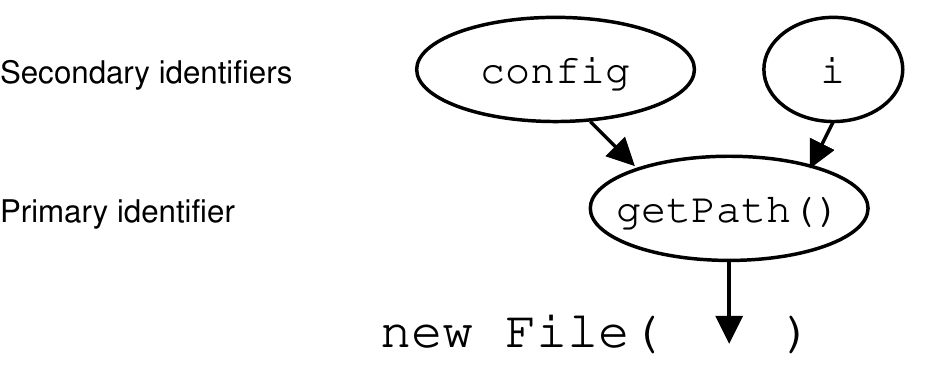}}
\caption{Dependency Graph of ``{\tt new File(config.getPath(i))}''
  and Its Primary and Secondary Identifiers}
\label{fig:argexpr}
\end{figure}

We first construct the data dependency graph of an expression by
traversing each term in its syntax tree.  For each term in the (sub-)
expression, the rules shown in Table \ref{tab:deprules} are applied
recursively. The obtained graph forms a lattice structure whose node
is either a variable access, method call, constant, or one of Java
operators. We then traverse the dependency graph from the top and
extract features at each node. Operator and constant nodes are
skipped. The most significant node that is close to the top is marked
as a primary identifier, and the second degree ones are marked as
secondary identifiers, and so on. As we move away from the top node in
the dependency graph, we obtain ternary or fourth-degree identifiers.
Fig. \ref{fig:argexpr} illustrates the primary and secondary
identifiers that appears in a method call
\verb+new File(config.getPath(i));+ The primary identifier of this
expression is \verb+getPath( )+. The secondary identifier is
\verb+config+ and \verb+i+.

Note that the chain of data dependency becomes longer as we obtain a
broader range of a dependency tree, i.e. the value represented at each
node has a more indirect influence to the entire expression.  For the
sake of simplicity, we discard fourth-degree or further identifiers.

\begin{table}[htbp]
\caption{Dependency Graph Rules}
\begin{center}
\begin{tabular}{|l|l|}
\hline
Expression & Dependency \\
\hline
    {\tt \#} (constant) &
    $\#$ \\
    {\tt A} (variable access) &
    $A$ \\
    {\tt A()} (method call) &
    $A()$ \\
    {\tt A.B} (field access) &
    $A \rightarrow B$ \\
    {\tt A.B()} (instance method call) &
    $A \rightarrow B()$ \\
    {\tt op A} (applying a unary operator) &
    $A \rightarrow \mathit{op}$ \\
    {\tt A op B} (applying a binary operator) &
    $A \rightarrow \mathit{op}, B \rightarrow \mathit{op}$ \\
    {\tt B = A} (assignment) &
    $A \rightarrow B$ \\
\hline
\end{tabular}
\end{center}
\label{tab:deprules}
\end{table}

\subsection{Word Segmentation}
\label{subsec:wordseg}

To give the prediction model more flexibility, we treat identifiers
not as a single feature but a set of features based on its tokens. For
example, ``{\tt getConfigPath}'' is segmented into three distinct
tokens: ``{\tt get}'', ``{\tt config}'' and ``{\tt path}''. Other than
tokenization, the classifier does not have any prior knowledge about
the natural language used in program identifiers.

We used a simple regex-based word tokenizer.  For a given string, we
search a longest substring that matches with {\tt ([A-Z][a-z]+|[A-Z]+)}
pattern. We chunk each matched substring as individual tokens.
Since the extent of each match is limited to consecutive
alphabets, both ``{\tt getConfigPath}'' and ``{\tt get\_config\_path}''
can be segmented to the same tokens. Each tokens is normalized to
lower case letters.

\subsection{ID3 Algorithm}
\label{subsec:id3}

\begin{figure}[htbp]
\centering
\fbox{\includegraphics[width=0.8 \linewidth]{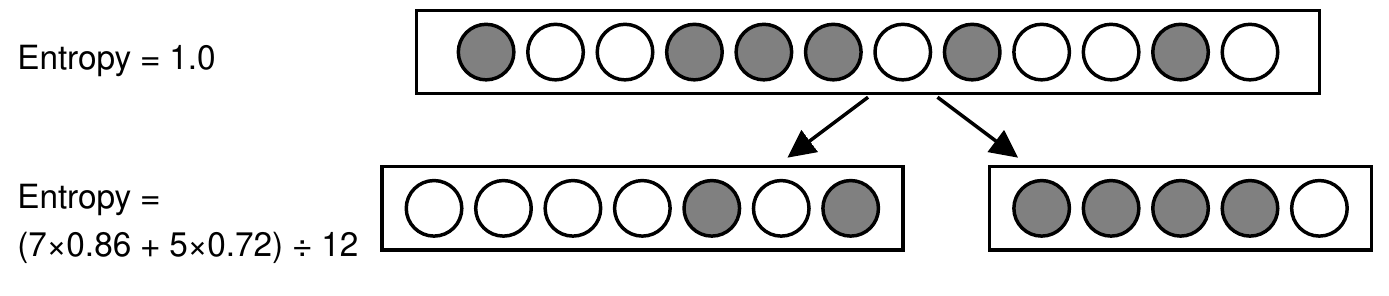}}
\caption{Splitting Tree with Minimal Average Entropy}
\label{fig:treesplit}
\end{figure}

ID3 is a recursive algorithm that produces an optimal decision tree in
terms of its total entropy.  Our ID3 implementation is fairly
straightforward. The way that the decision tree learner works is
following: it scans all the input instances and searches a test
that split the given instances the best. This means that a split with
the minimal average entropy is chosen (Fig. \ref{fig:treesplit}).
The average entropy of a split $S$ is calculated as:
\[ H_\mathit{avg}(S) = - \sum_{s_i \in S} \frac{s_i}{|S|} \log \frac{s_i}{|S|} \]
where $s_i$ is the number of equivalent items in the set.
The overall procedure of ID3 is shown in Fig. \ref{fig:id3algo}.

The algorithm starts with the most significant test, and then
repeatedly splits the subtrees until it meets a certain predefined
cutoff criteria; an important test tends to appear at the top of
the tree, and as it descends to its branches a less significant test
appears. In general, setting the cutoff threshold too small causes a
tree over-fitting problem, while setting it too large makes it
under-fitting. In our experiment, we found that setting the minimum
threshold to 10 instances produced the best results. The more detailed
mechanism is described in \cite{quinlan:93}.

Once the decision tree is built, it can be treated as a sequence of
if-then clauses. The classification process begins with the top node
of the tree; it performs a test at each branch and decides the
corresponding branch to descend. Each branch also has an associated
value (prediction). When it reaches at a leaf or there is no
corresponding branch, the process stops and the value associated with
the current branch is returned.

Table \ref{tab:id3feats} shows the list of ID3 features we used.
The test at each branch checks if a certain word is included in one of
the features.  Fig. \ref{fig:id3rules} shows an excerpt of
the obtained rules.

\begin{figure}[htbp]
\begin{Verbatim}[frame=single, fontsize=\scriptsize]
Features = [ ... ]
MinItems = 10

def buildTree(items):
    if len(items) < MinItems:
        default = getDefaultValue(items)
        return Leaf(default)
    else:
        bestSplit = None
        for f in Features:
            split = splitItemsByFeature(items, f)
            if calcEntropy(split) < calcEntropy(bestSplit):
                bestSplit = split
        nodes = []
        for s in split:
            nodes.append(buildTree(s))
        return Tree(nodes)
\end{Verbatim}
\caption{ID3 Algorithm (Python)}
\label{fig:id3algo}
\end{figure}

\begin{table}[htbp]
\caption{ID3 Features}
\begin{center}
\begin{tabular}{|l|l|}
\hline
Feature & Description \\
\hline
{\tt PrimaryFirstWords} & First Words of Primary Identifiers \\
{\tt PrimaryLastWords} & Last Words of Primary Identifiers \\
{\tt SecondaryFirstWords} & First Words of Secondary Identifiers \\
{\tt SecondaryLastWords} & Last Words of Secondary Identifiers \\
\hline
\end{tabular}
\end{center}
\label{tab:id3feats}
\end{table}

\begin{figure}[htbp]
\begin{Verbatim}[frame=single, fontsize=\scriptsize]
if "port" in PrimaryLastWords:
  if "get" in SecondaryFirstWords: ctype = PORT
  elif "host" not in PrimaryFirstWords: ctype = PORT
  ...
elif "height" in PrimaryLastWords:
  if "y" in PrimaryLastWords: ctype = YCOORD
  else: ctype = HEIGHT
elif "path" in PrimaryLastWords:
  if "host" in PrimaryLastWords:
    if PrimaryFirstWords == "host": ctype = HOST
    else: ctype = PATH
  elif "address" in PrimaryLastWords: ctype = OTHER
  ...
\end{Verbatim}
\caption{Obtained Rules (Python)}
\label{fig:id3rules}
\end{figure}

\subsection{Classification Results}
\label{subsec:results}

To measure the performance of our method, we conducted
leave-one-project-out cross validation; For each project, we use all
other 25 projects as the training data and use the one project as the
test data.  After repeating this project for 26 times, we took the
average of the precision and recall for each project across different
c-types. Table \ref{tab:id3results} shows the average precision and
recall as well as its F-score. The average F-score for all 12 c-types
was 83\%.

\begin{table}[htbp]
\caption{Classification Results for Each C-Types}
\begin{center}
\begin{tabular}{|l|r|r|r|}
\hline
C-Type & Precision & Recall & F-score \\
\hline
PATH    & 68.9\% & 91.8\% & 78.8\% \\
URL     & 61.3\% & 53.0\% & 56.8\% \\
SQL     & 70.4\% & 80.6\% & 75.2\% \\
HOST    & 70.0\% & 73.8\% & 71.8\% \\
PORT    & 84.6\% & 87.5\% & 86.0\% \\
XCOORD  & 95.7\% & 82.1\% & 88.3\% \\
YCOORD  & 97.5\% & 79.4\% & 87.5\% \\
WIDTH   & 92.0\% & 92.5\% & 92.2\% \\
HEIGHT  & 90.4\% & 93.4\% & 91.9\% \\
YEAR    & 100.0\% & 83.7\% & 91.1\% \\
MONTH   & 100.0\% & 77.0\% & 87.0\% \\
DAY     & 100.0\% & 61.1\% & 75.9\% \\
\hline
Average & 85.9\% & 79.6\% & 82.7\% \\
\hline
\end{tabular}
\end{center}
\label{tab:id3results}
\end{table}


\section{Discussions}
\label{sec:discussions}

As shown in Section \ref{subsec:results}, the average F-score of our
classifier was about 80\% for most c-types, except ``URL'' c-type, whose
F-score was less than 60\%. There are several reasons for this: First,
URL expressions tend to be long and has a number of components, as
shown in Table \ref{tab:topexprs}.  Most of these expressions are a
concatenation of multiple strings with {\tt +} operator, which is
exemplified in Table \ref{tab:topexprs}.  Also, since a URL typically
consists of a host name or path name, a URL expression tends to
include many PATH or HOST-associated expressions as its constituents,
which confuses the classifier. Indeed, this confusion is exhibited in
the confusion matrix shown in Table \ref{tab:confusion} \footnote{
This matrix is obtained by applying the classifier to its own training
set. Note that this is not to show the performance of the
classifier. Rather, it shows the limit of its discerning ability. }; a
lot of URL expressions were mistaken as PATH, HOST or PORT expressions.

Now, let us go back to our research questions:

\begin{enumerate}[leftmargin=*, label={RQ.\arabic*)}]
\item What kinds of high-level concepts do programmers commonly
  use in software projects?
\item How do programmers express such concepts in source code?
\item Is it possible to accurately predict such concepts from the
  source code appearance?
\end{enumerate}

First, we have observed that a different set of c-types appeared in
different projects as shown in Table \ref{tab:catcount}.  Unlike
general-purpose data types, the use of c-types depends on the domain of
the project. This somewhat agrees with our intuition; since c-types are
closer to application-specific types, its uses also depends on the
application domain.

The second and third questions are related. We have seen that a
decision tree-based classifier with simple features like Table
\ref{tab:id3feats} performed reasonably well for most c-types we
tested. This indicates the following: there are certain conventions
about how these c-types should be expressed and many programmers tend
to follow them. Therefore, for conceptual types that are as common and
well-defined as ours, it is relatively easy to identify them from the
surface features of the source code.  We think that our methodology
can be extended to a wider range of concepts in other third-party
libraries and frameworks.

\begin{table*}[htbp]
\caption{Confusion Matrix of C-Types}
\begin{center}
\begin{tabular}{|l|r|r|r|r|r|r|r|r|r|r|r|r|}
\hline
C-Type  & {\tiny PATH} & {\tiny URL} & {\tiny SQL} & {\tiny HOST} & {\tiny PORT} & {\tiny XCOORD} & {\tiny YCOORD} & {\tiny WIDTH} & {\tiny HEIGHT} & {\tiny YEAR} & {\tiny MONTH} & {\tiny DAY} \\
\hline
PATH  & 2274 &  28 &   1 &   2 &   3 &   0 &   0 &   0 &   0 &   0 &   0 &   0 \\
URL    &  82 & 737 &   2 &  17 &  54 &   0 &   0 &   0 &   0 &   0 &   0 &   0 \\
SQL    &   2 &   1 & 287 &   0 &   4 &   0 &   0 &   0 &   0 &   0 &   0 &   0 \\
HOST   &   7 &   5 &   0 & 424 &   2 &   0 &   0 &   0 &   0 &   0 &   0 &   0 \\
PORT   &   2 &   2 &   2 &   0 & 755 &   0 &   0 &   0 &   1 &   0 &   0 &   0 \\
XCOORD &   0 &   0 &   0 &   0 &   0 & 478 &   0 &   6 &   0 &   0 &   0 &   0 \\
YCOORD &   0 &   0 &   0 &   0 &   0 &   6 & 486 &   0 &   7 &   0 &   0 &   0 \\
WIDTH  &   0 &   0 &   0 &   0 &   0 &   3 &   0 & 597 &  10 &   0 &   0 &   0 \\
HEIGHT &   0 &   0 &   0 &   0 &   0 &   0 &   4 &  17 & 569 &   0 &   0 &   0 \\
YEAR   &   0 &   0 &   0 &   0 &   0 &   0 &   0 &   0 &   0 &  18 &   0 &   0 \\
MONTH  &   0 &   0 &   0 &   0 &   0 &   0 &   0 &   0 &   0 &   0 &  43 &   0 \\
DAY    &   0 &   0 &   0 &   0 &   0 &   0 &   0 &   0 &   0 &   0 &   0 &   7 \\
\hline
\end{tabular}
\end{center}
\label{tab:confusion}
\end{table*}

One of the possible ways to improve the classification accuracy is to
exploit a longer data flow between method calls and other statements. In
this paper, we treated individual method calls separately.  However,
when a method call is chained with another method call or statement,
we could take advantage of this additional restriction to further
refine the prediction result.

\subsection{Threats to Validity}
\label{subsec:threats}

Here we discuss the threats to validity of our findings:

\begin{itemize}
\item An incomplete method list. To extract a c-type expression,
  we need a list of API methods that specify the corresponding c-type.
  We manually searched the Java Standard API documentation to find
  the appropriate methods for each c-type, but we might have missed
  some methods.

\item Some c-types are rarely used in real world and we might not find
  enough examples. This is a classic data sparseness problem.
  Identifying some c-types might simply not be practical.

\item Open source selection bias. Our choice of the 26 open source
  projects might not be representative.

\item Some c-types cannot be well-defined. A primary example of this is
  the ``URL'' c-type. Technically, URL (Uniform Resource Locator) and
  URI (Universal Resource Identifier) are two different things
  \cite{uri-spec}.  URI is a broader concept which includes URL but
  can be used for offline entities such as book.  In this paper, we
  treated them interchangeably because these two concepts are almost
  identical in the context of network applications. However, certain
  c-type can be more confounding and we might not be able to
  distinguish them in a consistent way.  Another example would be
  ``file name'' and ``path name''. We are yet to know how many such
  c-types exist.

\item Potentially there are significantly more ``OTHER'' c-type
  expressions that we missed.  In this paper, we assigned a
  hypothetical ``OTHER'' c-type only to arguments in certain
  methods. However, this should not be limited to method calls only.
  If we are to identify the c-type of all expressions in a program,
  there will be many more OTHER expressions.  Training for all these
  OTHER expressions might confuse the classifier and end up with
  a much lower performance.

\end{itemize}


\section{Conclusion}

In this paper, we set out to examine how programmers express a
high-level concept such as path name or coordinates in source code.
We proposed a method to identify such concepts by using standard API
calls. We defined 12 c-types that are commonly used in many software
projects. Each c-type can be seen as an argument for the corresponding
API methods. We conducted experiments and obtained c-type expressions
from 26 open source projects. We constructed a decision tree-based
classifier that predicts the c-type from a given expression by
combining its lexical and data flow-centric features. We introduced
the notion of primary and secondary identifier. Our classifier
achieved 83\% average F-score for 12 c-types.

\section{Future Work}

There are several ways to extend our work.  A straightforward
extension is to support more c-types found in the Java Standard API or
other third-party APIs. Since a return value of API is typically also
well-defined, it is possible to extend the notion of c-type to return
values.

To improve the classification performance, one can take advantage of
more advanced data flow. For example, an inter-procedural data flow
between different functions or bidirectional data flow between
multiple statements can provide extra information to the classifier.
We could also use an advanced inference algorithm such as graph neural
network.


\bibliographystyle{IEEEtran}
\bibliography{../euske}

\end{document}